\documentclass[11pt,a4paper]{article}

\usepackage{pslatex}
\newcommand{\be}{\begin{equation}}
\newcommand{\ee}{\end{equation}}

\begin{document}

\title{Topology in the $SU(N_f)$ chiral symmetry restored phase of unquenched QCD and 
axion cosmology}

\author{Vicente ~Azcoiti \\
        Departamento de F{\'i}sica Te\'orica, Facultad de Ciencias \\
        Universidad de Zaragoza, Pedro Cerbuna 9, 50009 Zaragoza, Spain\\}

\maketitle

\begin{abstract}
We investigate the topological properties of unquenched $QCD$ on the basis of numerical 
results of simulations at fixed topological charge, recently reported by Borsanyi et al.. We demonstrate that 
their results for the mean value of the chiral condensate at fixed topological 
charge are 
inconsistent with the analytical prediction of the 
large volume expansion around the saddle point, and argue that the most plausible 
explanation for the failure of the saddle point  
expansion is a vacuum energy density $\theta$-independent at high temperatures, 
but surprisingly not too high $(T\sim 2T_c)$, a result which would imply a vanishing
topological susceptibility, and the absence of all physical effects of the $U(1)$ axial
anomaly at these temperatures. We also show that under a general assumption 
concerning the high temperature phase of
$QCD$, where the $SU(N_f)_A$ symmetry is restored, the analytical prediction for the 
chiral condensate at fixed topological charge is in very good agreement with the numerical 
results of Borsanyi et al., all effects of the axial anomaly should disappear, 
the topological susceptibility and all the $\theta$-derivatives of the vacuum 
energy density vanish and the theory becomes $\theta$-independent at any $T>T_c$ in the 
infinite volume limit.
\end{abstract}

\vfill\eject

\section{Introduction}

Understanding the role of the $\theta$ parameter in QCD and its connection with the strong CP 
problem is one of the major challenges for high energy theorists \cite{peccei}. 
The aim to elucidate the existence of new low-mass, weakly
interacting particles from a theoretical, phenomenological and experimental point of view is 
intimately related to this issue. 
The light particle that has gathered the most attention has been the axion, predicted by
Weinberg \cite{weinberg} and Wilczek \cite{wilczek} in the Peccei and Quinn mechanism 
\cite{pq} to explain the absence
of parity and temporal invariance violations induced by the QCD vacuum. The axion is one of 
the more interesting candidates to make the dark matter of the universe, and 
the axion potential, that determines the dynamics of the axion field, plays a fundamental 
role in this context. At high temperature, the potential can be calculated in the 
dilute instanton gas model \cite{turner}, but at medium and low temperatures 
interactions become 
non-perturbative, and a lattice QCD calculation is needed. 

The calculation of the 
topological susceptibility in QCD is already a challenge, but calculating the complete 
potential requires a strategy to deal with the so called sign problem, that is, 
the presence of a highly oscillating term in the path integral. 
Indeed Euclidean lattice gauge theory, our main non-perturbative tool for QCD studies 
from first principles, has not been able to help us much because of the imaginary 
contribution to 
the action coming from the $\theta$-term, that prevents the applicability of the importance 
sampling method \cite{vicari}. This is the main reason why the only progress in the 
analysis of the 
$\theta$-dependence of the vacuum energy density in pure gauge QCD, 
from first principles, 
reduces to the computation of the first few coefficients (up to order $\theta^6$) in the 
expansion of the free energy density in powers of $\theta$ \cite{pv}, \cite{bdpv},
\cite{new}, and the maximum temperature at which quenched simulations seem to give reliable 
results for the topological susceptibility is of the order of $1.5 T_c$ \cite{enrico}
\cite{japo}, \cite{javier_quen}, with $T_c$ 
the critical temperature for the 
chiral symmetry restoration phase transition. The situation in full 
QCD with dynamical fermions is, on the other hand, even worst \cite{martinelli}, 
\cite{petre}, \cite{javier}. 

The $QCD$ axion model relates the topological susceptibility $\chi_T$ at $\theta=0$ with 
the axion mass $m_a$ and decay constant $f_a$ through the relation $\chi_T = m^2_a f^2_a$. 
The axion mass is, on the other hand, an essential ingredient in the calculation of the 
axion abundance in the Universe. Therefore a precise computation of the temperature 
dependence of the topological susceptibility in $QCD$ becomes of primordial interest in 
this context. This is the reason why several calculations of the topological susceptibility 
in unquenched $QCD$ have been published in recent times \cite{martinelli},
\cite{petre}, \cite{javier}. 

The authors of reference \cite{martinelli} explore $N_f=2+1$ 
$QCD$ in a range of temperature going from $T_c$ to around $4T_c$, and their results for 
the topological susceptibility differ strongly, both in the size and in the temperature 
dependence, from the dilute instanton gas prediction, giving rise to a shift of the axion 
dark matter window of almost one order of magnitude with respect to the instanton 
computation. The authors of reference \cite{petre} observe however, in the same model, 
very distinct temperature dependences of the topological susceptibility in the ranges above 
and below 250 MeV: while for temperatures above 250 MeV, the dependence is found to be 
consistent with the dilute instanton gas approximation, at lower temperatures the fall-off of
topological susceptibility is milder. On the other hand a novel approach is proposed 
in reference 
\cite{javier}, the “fixed Q integration”, based on the computation of the mean value of the 
gauge action and chiral condensate at fixed topological charge $Q$, and they find a 
topological susceptibility many orders of magnitude smaller than that of 
reference \cite{martinelli} 
in the cosmologically relevant temperature region.

We want to show in this paper that an analysis of some of the numerical results reported 
in reference \cite{javier}, concerning the mean value of the 
chiral condensate at fixed topological 
charge, suggest that the vacuum energy density is $\theta$-independent at high temperatures, 
but surprisingly not too high $(T\sim 2T_c)$, a result which would imply a vanishing 
topological susceptibility, and the absence of all physical effects of the $U(1)$ axial 
anomaly at these temperatures. Since our analysis is based on the computation of physical 
quantities at fixed topological charge, we summarize some peculiar features of such a 
computation and derive an expression for the ratio of partition functions in
different topological sectors in section 2. In section 3 we show, provided that the vacuum 
energy density has a non-trivial $\theta$-dependence, that the difference of gauge actions 
and of chiral condensates between the $Q$ and vanishing topological sectors are of the 
order of the 
inverse lattice volume $\frac{1}{V_xL_t}$, and proportional to the square of the 
topological charge Q, in both cases. In this section we also compare our analytical results with 
the numerical results reported in reference \cite{javier}. 
The absence of the typical effects of the 
$U(1)_A$ anomaly in the chiral symmetry restored phase of $QCD$ at high-temperature was 
suggested years ago \cite{cohen1}, \cite{cohen2}, and investigated later on in 
\cite{cuatro}, \cite{cinco}, \cite{seis}, \cite{siete}, \cite{ocho}, \cite{nueve},
\cite{diez}, \cite{once}, \cite{doce}, \cite{docebis}. 
In section 4 we show, under very general assumptions, 
that all effects of the axial anomaly should disappear in the high temperature phase of 
$QCD$, where the $SU(N_f)_A$ symmetry is restored. The topological susceptibility and all 
the $\theta$-derivatives of the vacuum energy density should vanish and the theory should 
become $\theta$-independent. Our conclusions are reported in section 5.

\section{QCD with $\theta$-term}

Quantum Field Theories with a topological term in the action are a subject of
interest in high energy particle physics and in solid state physics. In
particle physics, these models describe particle interactions with a $CP$
violating term. The inclusion of this term in the $QCD$ Lagrangian was the
result of the discovery of the $U(1)$ axial anomaly, which solved the 
$U(1)_A$ problem but generated a new problem, the strong $CP$ problem.

The Euclidean continuum Lagrangian of $N_f$ flavors $QCD$ with a $\theta$-term is 

\begin{equation}
L = \sum_f L^f_F + \frac{1}{4} F^a_{\mu\nu}\left(x\right)F^a_{\mu\nu}\left(x\right) 
+ i\theta \frac{g^2}{64\pi^2} \epsilon_{\mu\nu\rho\sigma} 
F^a_{\mu\nu}\left(x\right)F^a_{\rho\sigma}\left(x\right)
\label{eulagran}
\end{equation}
with $L^f_F$ the fermion Lagrangian for the $f$-flavor, and 

\begin{equation}
Q = \frac{g^2}{64\pi^2} \int d^4x\epsilon_{\mu\nu\rho\sigma}
F^a_{\mu\nu}\left(x\right)F^a_{\rho\sigma}\left(x\right)
\label{ftopcharg}
\end{equation}
is the topological charge of the gauge configuration, which takes integer values.

In this section I want to summarize a few interesting features, some of them well known, of 
$QCD$ with a topological term in the action. One of these features concerns the fact that 
the mean value of any intensive operator in $QCD$ at $\theta=0$ can be computed in any 
fixed topological sector \cite{wolfgang}, \cite{aokiqfix}, in particular 
in the $Q=0$ 
topological sector. Even if this result 
seems a paradox, since the zero charge topological sector is free from the $U(1)_A$ anomaly
and breaks spontaneously chiral symmetry, 
I will show how one can re-conciliate the absence of the $U(1)_A$ anomaly with a finite 
non-vanishing mass 
for the $\eta$ meson. I will discuss separately the one-flavor model and the case of 
several flavors, and will derive a expression for the ratio of partition functions in 
different topological sectors, which will be useful in the next section.

\subsection{The one flavor model}

Concerning the one flavor model, where the only axial symmetry is an
anomalous $U(1)$ symmetry, the standard wisdom on the vacuum structure of
this model in the chiral limit is that it is unique at each given value
of $\theta$, the $\theta$-vacuum. Indeed, the only plausible reason to
have a degenerate vacuum in the chiral limit would be the spontaneous
breakdown of chiral symmetry, but since it is anomalous, actually there is
no symmetry. Furthermore in contrast to what happens
when chiral symmetry is spontaneously broken, the infinite volume limit and the chiral limit 
commute. In fact, due to the chiral anomaly, the model shows a mass gap in the chiral limit 
and therefore all correlation lengths are finite in physical units.

An elegant realization of all these ideas is the Leutwyler and
Smigla (L-S) approach \cite{Smilga}.
This approach is based, for the one-flavor model, on the assumption
that the vacuum energy or free energy density can be expanded in powers of the
fermion mass m, treating the quark mass term as a perturbation. Indeed, as previously stated, 
the spectrum of the one-flavor model, due to the chiral anomaly, does not contain
massless particles and therefore the perturbation series in powers of
the fermion mass $m$ should not give rise to infrared divergences. This
expansion will be then an ordinary Taylor series

\begin{equation}
-E (m,\theta) = -E_0 + \Sigma m cos\theta +O(m^2),
\label{LS}
\end{equation}
giving rise to the following expansions for the scalar and pseudoscalar
condensates

\begin{equation}
\left\langle\bar u u\right\rangle =  \Sigma cos\theta +O(m)
\end{equation}

\begin{equation}
\left\langle i\bar u\gamma_5 u\right\rangle =  \Sigma sin\theta +O(m)
\label{LSc}
\end{equation}
The resolution of the U(1) axial problem is obvious in this
approach. Indeed the expression for the free energy density (\ref{LS}) tell us that
the topological susceptibility $\chi_T$ has the following expansion

\begin{equation}
\chi_T = \Sigma m cos\theta +O(m^2)
\end{equation}
and then the divergence in the chiral limit of the first term in the
right-hand side of the equation relating the pseudoscalar susceptibility
 
$$\chi_p = \int\left\langle i\bar u(x)\gamma_5 u(x) i\bar u(0)\gamma_5 u(0)
\right\rangle d^4 x$$
with the chiral condensate and $\chi_T$

\begin{equation}
\chi_p = \frac{\left\langle\bar u u\right\rangle}{m} -
\frac{\chi_T}{m^2}\,
\label{Trans-II}
\end{equation}
is compensated with the divergence of second term in this equation, giving
rise to a finite pseudoscalar susceptibility or equivalently a finite mass for
the $\bar u\gamma_5 u$ meson.

All these features can be understood in simple words. Due to the chiral
anomaly a non-vanishing value of the chiral condensate does not break any symmetry. The
Goldstone theorem is not fulfilled because there is no spontaneous symmetry
breaking.

The (L-S) formalism was developed in the continuum. However, there is a
lattice regularization, the Ginsparg-Wilson (G-W) fermions \cite{Ginsparg}
from which the overlap fermions \cite{Neuberger} are an explicit realization,
which shares with the continuum all essential ingredients and gives at the
same time mathematical rigor to all developments. Indeed G-W fermions have a
$U(1)$ anomalous symmetry \cite{Luscher}, good chiral properties, a
quantized topological charge, and allow us to establish and exact index
theorem on the lattice \cite{Victor}. Furthermore G-W fermions, contrary to Wilson fermions, 
are free from phases where parity an flavor symmetries are spontaneously 
broken \cite{monos}. We will use this lattice regularization in what follows.

With this in mind we can write 
for the free energy density $E$, scalar $S$ and pseudoscalar $P$
condensates, pseudoscalar susceptibility $\chi_p$ and topological
susceptibility $\chi_T$ the same expressions as in the L-S approach,

\begin{equation}
-E (\beta,m,\theta) = -E_{0}\left(\beta\right) + \Sigma m cos\theta +O(m^2),
\label{free}
\end{equation}

\begin{equation}
\left\langle S\right\rangle =  \Sigma cos\theta +O(m)
\label{scondensate}
\end{equation}

\begin{equation}
\left\langle P\right\rangle =  \Sigma sin\theta +O(m)
\label{pcondensate}
\end{equation}

\begin{equation}
\chi_T =  \Sigma m cos\theta +O(m^2)
\label{tops}
\end{equation}

\begin{equation}
\chi_p = \frac{\left\langle S\right\rangle}{m} -
\frac{\chi_T}{m^2}\,
\label{trans}
\end{equation}
where these expressions are now valid at finite lattice spacing $a$ and
finite lattice volume $V$, and $\Sigma$ depends on both parameters with a
finite non-vanishing value in the infinite volume limit. $\beta$ is the inverse gauge 
coupling and we omit the $\beta$-dependence of $\Sigma$ for simplicity. 

The partition function of the model can be written as a sum over all topological sectors $Q$ 
of the partition function in each topological sector times a $\theta$-phase factor as follows

\begin{equation}
Z(\theta) = \sum_{Q} Z_Q e^{i\theta Q} 
\label{zeta}
\end{equation}
where Q takes all integer values, and it is bounded at finite volume by the number of 
degrees of freedom.

At large lattice volume $V$ the partition function should behave as

\begin{equation}
Z\left(\theta\right) = e^{-V E\left(\beta,m,\theta\right)}
\label{zetalarge}
\end{equation}
with $E\left(\beta,m,\theta\right)$ given by (\ref{free}). On the other hand the mean value
of any intensive operator $O$, as for instance the scalar and pseudoscalar
condensates or any correlation function, in a given topological sector $Q$, can be 
computed as follows

\begin{equation}
\left< O\right>_{Q} = {{\int d\theta \left< O\right>_\theta Z(\theta, m)}
e^{-i \theta Q}\over {\int d\theta Z(\theta,m)e^{-i \theta Q}}}.
\label{mascurioso}
\end{equation}

Since the vacuum energy density, as a function of $\theta$, has its absolute minimum at 
$\theta=0$, equations (\ref{free}) and (\ref{mascurioso}) tell us that the mean value of any 
intensive operator at $\theta=0$ and non-vanishing fermion mass can be computed in any 
fixed topological sector. Indeed, equation (\ref{mascurioso}) gives in the infinite lattice 
volume limit the following relation

\begin{equation}
\left< O\right>_{Q} = \left< O\right>_{\theta=0}
\label{mascuriosob}
\end{equation}

We can apply equation (\ref{mascuriosob}) to the computation of the pseudoscalar correlation 
function $<P(x) P(0)>_{\theta=0}$ by computing it in the vanishing charge topological sector. 
But this sector is anomaly free, and breaks spontaneously chiral symmetry in order to give 
a non-vanishing value $\Sigma$ for the chiral condensate in the chiral limit. The 
pseudoscalar meson susceptibility diverges and the Goldstone theorem should tell us that 
the pseudoscalar meson is massless in the chiral limit. The loophole in this argument is 
that in systems with a global 
constraint, the divergence of the susceptibility does not necessarily implies
a divergent correlation length. Indeed the susceptibility must be computed by integrating out 
the correlation function over all distances, and then taking the infinite
volume limit, in this order. In systems with a global constraint, the infinite volume limit 
and the space-integral of the correlation function do not necessarily commute. A very simple 
example of that is the Ising model at infinite temperature with an even number of spins
and with vanishing full magnetization as global constraint. In such a case
one has for the spin-spin correlation function

$$\left< s_i^2 \right>=1$$
$$\left< s_i s_j \right>= -{{1}\over {\left(V-1\right)}}$$
The integral of the infinite volume limit of the correlation function is equal to 1,
whereas the infinite volume limit of the integrated correlation function vanishes.
The correlation function has a contribution of order $1/V$ that violates cluster at 
finite volume and vanishes in
the infinite volume limit, but that gives a finite contribution to the
integrated correlation function.
This example, even if very simple, is illustrative because this is in fact what happens 
for the pseudoscalar correlation function.

Coming back to QCD with a $\theta$-term, the standard wisdom on this model is that it 
has no phase transition at $\theta=0$. Then we can expand the pseudoscalar correlation 
function in powers of the $\theta$ angle 
as follows

\begin{equation}
<P(x) P(0)>_{\theta} = <P(x) P(0)>_{\theta=0} + h(x, m_u) \theta^2 
+ O(\theta^4)
\label{expansion}
\end{equation}
where

\begin{equation}
h(x, m_u) = \left\langle S(x) S(0)\right\rangle_{\theta=0} - 
\left\langle P(x) P(0)\right\rangle_{\theta=0} + O(m_u)
\label{ache}
\end{equation}
The vacuum energy density (\ref{free}) can also be expanded in powers of $\theta$ as 

\begin{equation}
-E \left(\beta,m_u,\theta\right) = -E_{0}\left(\beta,m_u\right) 
- \frac{1} {2} \chi_T\left(\beta,m_u\right)\theta^2 +O(\theta^4)
\label{expansion2}
\end{equation}
with 

\begin{equation}
\chi_T\left(\beta,m_u\right) = m_u \Sigma + O(m_u^2)
\label{expansion3}
\end{equation}

Taking into account equations (\ref{mascuriosob}) and (\ref{expansion}-\ref{expansion2}) and 
making an expansion around the saddle point solution we can write the following equation 
for the pseudoscalar correlation function in the zero-charge topological sector

$$<P(x) P(0)>_{Q=0} = <P(x) P(0)>_{\theta=0} +$$ 
\begin{equation}
\frac{1}{V} \frac{\left\langle S(x) S(0)\right\rangle_{\theta=0} -
\left\langle P(x) P(0)\right\rangle_{\theta=0}+ O(m_u)}{\chi_T} + 
+ O\left(\frac{1}{V^2}\right)
\label{saddlepoint}
\end{equation}

Equation (\ref{saddlepoint}) shows, as in the simple Ising model case, a violation of 
cluster at finite volume for the pseudoscalar correlation function in the zero-charge 
topological sector. In the infinite volume limit, the pseudoscalar correlation function 
in the zero-charge topological sector and in $QCD$ at $\theta=0$ agree, as expected. 
Concerning susceptibilities we can write, by integrating out equation (\ref{saddlepoint}) 
and taking the infinite volume limit the following relation

\begin{equation}
\chi_{p,Q=0} = \chi_p + \frac{\Sigma^2 + O\left(m_u\right)}{\chi_T}
\label{susceptibilities}
\end{equation}
where we have made use of the fact that in the infinite volume limit intensive operators 
do not fluctuate 

$$\left\langle \left(\frac{1}{V}\sum_x  S(x)\right)^2\right\rangle = 
\left\langle \frac{1}{V}\sum_x S(x)\right\rangle^2$$

$$\left\langle \left(\frac{1}{V}\sum_x P(x)\right)^2\right\rangle =
\left\langle \frac{1}{V}\sum_x P(x)\right\rangle^2$$

The dominant contribution of the second term in equation (\ref{susceptibilities}) in the 
chiral limit diverges with the quark mass as $\Sigma/m_u$, whereas 

$$\chi_{p,Q=0} = \frac{\left\langle S(x)\right\rangle}{m_u}$$
Combining these results we get, notwithstanding the pseudoscalar susceptibility 
diverges in the zero charge sector in the chiral limit, that the pseudoscalar 
susceptibility in one-flavor $QCD$ is finite and the pseudoscalar meson is massive.
The pseudoscalar susceptibility in the $Q=0$ sector diverges in the chiral limit 
not because of a divergent 
correlation length but as a consequence of the cluster violating contributions 
to the pseudoscalar correlation function (\ref{saddlepoint}), which are singular at 
$m=0$ and of order $\frac{1}{V}$, and 
which give a finite singular contribution to $\chi_{p,Q=0}$.

To conclude the discussion on the one-flavor model, we want to remark that the validity of 
the commutation of the infinite volume limit and the chiral limit in this model does not 
apply to the zero charge topological sector. Indeed, as previously stated, the zero charge 
topological sector breaks spontaneously chiral symmetry, and even if all correlation lengths 
are finite in this sector, there are divergent susceptibilities in the chiral limit. 
We have seen that the 
pseudoscalar susceptibility diverges in this sector, but also the scalar susceptibility 
$\chi_s$ at vanishing quark mass can be computed as 

\begin{equation}
\chi_{s,Q=0,m_u=0} = \frac{1}{2} \left(\chi_{s,m_u=0} + \chi_{p,m_u=0}\right) + 
\frac{V}{2}\Sigma^2
\label{susceptibilitiesb}
\end{equation}
which shows explicitly the divergence with the lattice volume $V$, and makes the perturbative 
expansion of the chiral condensate in powers of $m_u$ ill-defined in the infinite 
volume limit.

\subsection{Several flavors}

QCD with several flavors shows some important differences with respect to the one flavor 
case. The model also suffers from the chiral anomaly but has a spontaneously broken 
$SU(N_f)$ chiral symmetry in the chiral limit at any temperature below the critical 
temperature of the chiral transition $T_c$. There are divergent correlation lengths for 
$T<T_c$ in this limit and, contrary to the one flavor case, the infinite volume limit and 
the chiral limit do not 
commute if $T<T_c$. However the essential features previously described for the one-flavor 
model still work in the several flavors case. Equation (\ref{trans}) reads now 

\begin{equation}
\chi_p = \frac{\left\langle S\right\rangle}{m} -
N_f^2\frac{\chi_T}{m^2}\,
\label{transsf}
\end{equation}
where $\chi_p$ stands now for the flavor singlet pseudoscalar susceptibility and $S$ is 
the flavor singlet scalar condensate. The vacuum energy density can also be expanded in 
this case in powers of the $\theta$-angle as

\begin{equation}
E (\beta,m_f,\theta) = E_{0} - \frac{1} {2} \chi_T\left(\beta,m_f\right)\theta^2 +O(\theta^4)
\label{expansion2s}
\end{equation}
and equations (\ref{zeta}), (\ref{zetalarge}), (\ref{mascurioso}) and (\ref{mascuriosob}) 
also work for several flavors.

Let us write the expression, that we will use in the following, for the 
ratio of the partition functions in the 
$Q$ topological sector $Z_Q$ and in the vanishing topological sector $Z_0$

\begin{equation}
\frac{Z_Q}{Z_0} = {{\int d\theta e^{-i Q \theta} Z(\theta, m)}
\over {\int d\theta Z(\theta,m)}}
\label{ratio}
\end{equation}
and its expansion around the saddle point solution

\begin{equation}
\frac{Z_Q}{Z_0} = 1 - \frac{1}{V_x L_t} \frac{Q^2}{2\chi_T} + O\left(\frac{1}{V^2}\right)
\label{zrexpansion}
\end{equation}
where $V_x$ is the spatial lattice volume and $L_t$ the number of lattice points in time 
direction.
Equation (\ref{zrexpansion}) implies that all topological sectors have the same probability 
in the infinite spatial volume limit at any temperature $T= 1/L_t$. Otherwise 
the saddle point expansion breaks down, the most plausible reason for that being that 
the vacuum energy density (\ref{expansion2s}) be $\theta$-independent.

\section{The finite temperature chiral transition}

We want to explore in this section the physical consequences of equation (\ref{zrexpansion}) 
on the temperature dependence of the topological susceptibility. 

Taking the logarithm in (\ref{zrexpansion}) we get

\begin{equation}
\log \frac{Z_Q}{Z_0} = - \frac{1}{V_x L_t} \frac{Q^2}{2\chi_T} 
+ O\left(\frac{1}{V^2}\right)
\label{logzrexpansion}
\end{equation}
and the following expressions for the logarithmic derivatives respect to the inverse 
square gauge coupling $\beta$ and fermion masses $m_f$

\begin{equation}
\left\langle S_g\right\rangle_Q - \left\langle S_g\right\rangle_{Q=0} = 
\frac{Q^2}{V_x L_t}\frac{1}{2\chi^2_T}\frac{\partial\chi_T}{\partial\beta}
+ O\left(\frac{1}{V^2}\right)
\label{difactions}
\end{equation}

\begin{equation}
\left\langle \sum_x S_f\left(x\right)\right\rangle_Q -
\left\langle \sum_x S_f\left(x\right)\right\rangle_{Q=0} = 
\frac{Q^2}{V_x L_t}\frac{1}{2\chi^2_T}\frac{\partial\chi_T}{\partial m_f}
+ O\left(\frac{1}{V^2}\right)
\label{difcondensates}
\end{equation}
where $S_g$ and $S_f\left(x\right)$ in (\ref{difactions}) are the lattice pure gauge 
action and the scalar chiral condensate respectively.

There are two remarkable properties of the difference of gauge actions and of 
chiral condensates between the $Q$ and vanishing topological sectors in (\ref{difactions}), 
(\ref{difcondensates}): they are of the order of the inverse lattice volume 
$\frac{1}{V_x L_t}$ and proportional to the square of the topological charge 
$Q$ in both cases.

The numerical results for 
$\left\langle S_g\right\rangle_Q - \left\langle S_g\right\rangle_{Q=0}$ and 
$\left\langle \sum_x S_f\left(x\right)\right\rangle_Q -
\left\langle \sum_x S_f\left(x\right)\right\rangle_{Q=0}$ 
reported in reference \cite{javier} show 
a finite non-vanishing contribution in the infinite volume 
limit, linear in $|Q|$ for both quantities. These results 
have been obtained 
from numerical simulations of lattice QCD with $N_f=3+1$ staggered dynamical quarks at 
$T\sim 5T_c$ and $N_f=2+1$ overlap fermions, the last in a range of temperatures running 
from $2T_c$ to $4T_c$ \cite{javier}, and also in the quenched model \cite{frison}. 
The numerical results of reference \cite{javier} show furthermore a value of 
$\left\langle \sum_x S_f\left(x\right)\right\rangle_{Q=1} - 
\left\langle \sum_x S_f\left(x\right)\right\rangle_{Q=0}$,

\begin{equation}
\left\langle \sum_x S_f\left(x\right)\right\rangle_{Q=1} - 
\left\langle \sum_x S_f\left(x\right)\right\rangle_{Q=0} \approx \frac{1}{m_f}
\label{ascondensate}
\end{equation}
independent of the temperature, in the range of temperatures reported 
($300 MeV - 650 MeV$). 

Summarizing, the results reported in \cite{javier} for the difference of the 
gauge action and of the chiral condensate between the $Q$ and vanishing 
topological sectors, obtained from numerical simulations of $QCD$ at $T>T_c$, are 
in contradiction with the corresponding results obtained from the expansion 
around the saddle point (\ref{difactions}), (\ref{difcondensates}), which should 
hold in the large volume limit.

There are only two plausible explanations for such a contradiction:

\begin{itemize}
\item
The results of reference \cite{javier} are afflicted from strong volume 
corrections.

\item
The saddle point expansion fails to reproduce the correct behavior of physical 
quantities in the large volume limit
\end{itemize}
Since the authors of reference \cite{javier} exclude, from their numerical analysis 
of the difference of the gauge action and of the chiral condensate between the $Q$ 
and vanishing topological sectors, large finite size corrections to these 
quantities (see Figures S19 and S22 of \cite{javier}), the only plausible 
explanation is the failure of the expansion around the saddle point. 
But the only reason for the failure of the saddle point expansion is that the 
main ingredient of this expansion, the fact that $Z(\theta)$ defines in (\ref{ratio}) an 
integration measure extremely sharped around $\theta=0$, does not work. Since $Z(\theta) = 
e^{-V_xL_t E(\beta, m_f, \theta)}$ and $E(\beta, m_f, \theta)$ has its absolute maximum at 
$\theta=0$ for any non-vanishing value of the fermion mass $m_f$, and gauge coupling $\beta$, 
we should conclude, as previously stated, that the vacuum energy density is 
$\theta$-independent at high temperatures, but surprisingly not too high ($T\sim 2T_c$).
A $\theta$-independent vacuum energy density for physical temperatures above a given 
temperature $T_{ch}$ would imply a vanishing topological susceptibility, and the absence of 
all physical effects of the $U(1)$ axial anomaly at these temperatures. 

Years ago Thomas Cohen \cite{cohen1}, \cite{cohen2} showed, 
assuming the absence of the zero mode's 
contribution, that all the disconnected contributions to 
the two-point correlation functions in the $SU(2)_A$ chiral symmetry restored phase 
at high-temperature vanish in the chiral limit. The main conclusion of this work was that 
the eight scalar and pseudoscalar mesons $\sigma, {\bf \bar\pi}, \eta, {\bf \bar\rho}$, 
should have the same mass 
in the chiral limit, the typical effects of the $U(1)_A$ anomaly being absent in this phase. 
This issue has been investigated both analytically and from numerical simulations in 
\cite{cuatro}, \cite{cinco}, \cite{seis}, \cite{siete}, \cite{ocho}, \cite{nueve}, 
\cite{diez}, \cite{once}, \cite{doce}, \cite{docebis}. 
In particular Sinya Aoki and collaborators have 
reported numerical results 
from simulations of $QCD$ with overlap fermions \cite{nueve}, \cite{once} which show a 
degeneracy of 
the ${\bf\bar\pi}$ and $\eta$ correlators. In addition 
they have also shown in \cite{diez}, by 
studying multi-point correlation functions in various channels, that the $U(1)_A$ anomaly 
becomes invisible in susceptibilities of scalar and pseudo-scalar mesons in the $SU(2)_A$ 
chiral symmetric phase of $QCD$ with two overlap quarks. 

In the next section I will argue that the effects of the chiral anomaly on the 
meson spectrum, and in any physical observable, should disappear in the 
high temperature $SU(N_f)_A$ chiral symmetric phase of $QCD$. 

\section{The restoration of the $U(1)_A$ symmetry at any $T>T_c$}

We want to show in this section, on very general grounds, how all effects of the axial 
anomaly should disappear in the high temperature phase of $QCD$, where the 
$SU(N_f)_A$ symmetry is restored. 
The topological susceptibility and all the $\theta$-derivatives of the vacuum energy density 
should vanish and the theory should become $\theta$-independent.

The only general assumption of this section is that in the high temperature phase of $QCD$, 
where the $SU(N_f)_A$ symmetry is restored, the spectrum shows a mass gap even in the chiral 
limit. All correlation 
lengths are finite in physical units, none symmetry is spontaneously broken, the model 
is free from infrared divergences, and the 
perturbative expansion of the vacuum energy density and of the chiral condensate in 
powers of the quark mass converges for 
every $\theta$ (phase transitions in $\theta$ are not expected 
at $T>T_c$ \cite{delia}, \cite{new}). A 
finite spatial lattice volume of linear size much larger than the inverse mass gap should 
be enough to reproduce the correct physical results, and contrary to what happens in the 
low temperature broken phase, the infinite volume limit and the chiral limit should commute. 
The situation is similar to that of the one-flavor model previously discussed, where the 
chiral anomaly, and therefore the absence of spontaneous chiral symmetry breaking, was the 
responsible for the mass gap in the spectrum of this model. However, contrary to the 
one-flavor case, the zero charge topological sector does not show spontaneous symmetry 
breaking, and all susceptibilities are finite in the chiral limit in this sector. 
This suggests that the 
validity of the perturbative expansion in powers of the quark mass $m$, and of the 
commutation of the infinite volume limit and the chiral limit, applies also to this sector, 
and we will make use of this in what follows.

We will first discuss the two-flavor case and will comment on the 
extension 
of the results to $N_f\ge3$. We will also show that equation (\ref{ascondensate}) holds
in the chiral symmetry restored phase, up 
to order $m_f$ corrections, a result which, as previously stated, has been observed in the 
numerical simulations reported in \cite{javier} at $T=2T_c$. As always along this paper, 
we work in a lattice with a fermion regularization that, as the overlap fermions, obey the 
Ginsparg-Wilson relation.

\subsection{The two flavor model}

To fix the notation let be $S(x) = S_u(x) + S_d(x)$ and $P(x) = P_u(x) + P_d(x)$ 
the sum of the up and down scalar and pseudoscalar condensates respectively, and 
$\chi_{s,m=0,V}$ and $\chi_{p,m=0,V}$ 
the flavor singlet scalar and pseudoscalar susceptibilities at $m=m_u=m_d=0$, and finite 
lattice volume $V=L^3_s L_t$. Taking into account that $S(x)$ and $P(x)$ transform like 
a vector under $U_A(1)$ chiral anomalous rotations we can write for the expansion of the 
mean value of the chiral condensate in powers of $m$

\begin{equation}
\left\langle S\left(x\right) \right\rangle_\theta = \chi_{s,m=0,V} m - 
\sin^2 \frac{\theta}{2}\left(\chi_{s,m=0,V} - \chi_{p,m=0,V}\right) m
+ O\left(m^3\right)
\label{scondensate2f}
\end{equation}
which gives the following expression for the vacuum energy density

$$
-E_V \left(\beta,m,\theta\right) = -E_{0,V}\left(\beta\right) + 
\frac{1}{2} \chi_{s,m=0,V} m^2 - \frac{1}{2} \sin^2 
\frac{\theta}{2} \left(\chi_{s,m=0,V}-\chi_{p,m=0,V}\right) m^2 
$$
\begin{equation}
\hskip -5.5cm + O\left(m^4\right) 
\label{evac2f}
\end{equation}
where $E_{0,V}(\beta)$ is the vacuum energy density at $m=0$, which depends only on the 
inverse gauge coupling $\beta$.

Equation (\ref{evac2f}) gives for the topological susceptibility at $\theta=0$ the following 
relation with the scalar and pseudoscalar flavor-singlet susceptibilities

\begin{equation}
\chi_{T,V} = \frac{m^2}{4} \left(\chi_{s,m=0,V}-\chi_{p,m=0,V}\right) + + O\left(m^4\right)
\label{tops2f}
\end{equation}
which is of the order of $m^2$, as expected.

Equations (\ref{scondensate2f}) and (\ref{evac2f}) allow us to write the following expansion 
in powers of $m$ for the mean value of the chiral condensate in the $Q=0$ topological sector

\begin{equation}
\left\langle S\left(x\right) \right\rangle_{Q=0} = \chi_{s,m=0,V} m -
\frac{1}{2}\left(\chi_{s,m=0,V} - \chi_{p,m=0,V}\right) m
+ O\left(m^3\right)
\label{scondensate2fq0}
\end{equation}
As previously discussed, the large lattice volume expansion around the saddle point 
predicts, provided that the vacuum energy density shows a non-trivial $\theta$-dependence,  
that the chiral condensate in any fixed topological sector equals the chiral condensate in 
the full theory at $\theta=0$, in the large volume limit, up to of the order of $\frac{1}{V}$ 
corrections. Then the only way to keep the validity
of the expansion of the chiral condensate and the vacuum energy density in powers of the
quark mass $m$ is that $\chi_{s,m=0,V} - \chi_{p,m=0,V}$ is $O\left(\frac{1}{V}\right)$

\begin{equation}
\chi_{s,m=0,V} - \chi_{p,m=0,V} \sim O\left(\frac{1}{V}\right)
\label{solucion}
\end{equation}

Equation (\ref{solucion}) implies that the topological susceptibility (\ref{tops2f}) 
vanishes, the 
scalar and pseudoscalar susceptibilities are equal in the chiral limit, and therefore the 
eight scalar and pseudoscalar mesons $\sigma, {\bf \bar\pi}, \eta, {\bf \bar\rho}$,
should have the same mass in this limit.

The analysis here performed can be extended to higher orders in the expansions 
(\ref{scondensate2f})-(\ref{scondensate2fq0}), 
getting as a result new conditions, analogous to (\ref{solucion}), which show 
that the theory should 
be $\theta$-independent in the infinite volume limit and that all the effects of the chiral 
anomaly are missed.

There is a simpler way to understand all these features. The  
vacuum energy density can be parameterized as follows

\begin{equation}
E_V \left(\beta,m,\theta\right) -E_V \left(\beta,m,0\right) = 
m^2 \theta^2 f\left(\beta,m,\theta^2\right)
\label{evac2fsubs}
\end{equation}
with $f\left(\beta,m,\theta^2\right)>0$ for every $\theta\in(-\pi,\pi]$, since $\theta=0$ is 
assumed to be the only absolute minimum of the vacuum energy density. It 
is an even function of $\theta$ ($f\left(\beta,m,\theta^2\right)$ is also an even function of 
$m$ in the two-flavor model) that vanishes at $m=0$. The subtracted full chiral condensate 
$\left\langle \sum_x S\left(x\right) \right\rangle_{\theta=0} -
\left\langle \sum_x S\left(x\right) \right\rangle_{Q=0}$ is on the other hand finite 
in the infinite volume limit, and can be computed as follows

$$
\hskip -6cm
\left\langle \sum_x S\left(x\right) \right\rangle_{\theta=0} - 
\left\langle \sum_x S\left(x\right) \right\rangle_{Q=0} = 
$$
\begin{equation}
V_x L_t\frac{\int d\theta \left(2m f\left(\beta,m,\theta^2\right) + 
m^2 \partial_m f\left(\beta,m,\theta^2\right)\right) 
\theta^2
e^{-V_x L_t m^2 \theta^2 f\left(\beta,m,\theta^2\right)}}{\int 
d\theta e^{-V_x L_t m^2 \theta^2 f\left(\beta,m,\theta^2\right)}}
\label{scondensate2fsubs}
\end{equation}
which obviously vanishes at $m=0$. 

We can compute the subtracted full chiral condensate at any non-vanishing quark mass 
$m$ by doing the expansion of (\ref{scondensate2fsubs}) around the saddle point, and the 
final result for the dominant contribution in the $m\rightarrow 0$ limit is

\begin{equation}
\left\langle \sum_x S\left(x\right) \right\rangle_{\theta=0} - 
\left\langle \sum_x S\left(x\right) \right\rangle_{Q=0} =
\frac{1}{m}
\label{scondensate2fsubsdom}
\end{equation}
Then, if we want to keep the validity of the expansion of the vacuum energy density in 
powers of the quark mass $m$, and of the commutation 
of the infinite volume limit and the chiral limit, we need to invalidate the saddle point 
expansion, and this requires that the $\theta$-dependent part of the vacuum energy density 
(\ref{evac2fsubs}) be at least of the order of $\frac{1}{V}$.

We can also compute the mean value of the chiral condensate in the $Q=1$ topological sector 
under this condition (or condition (\ref{solucion})). The final result is

\begin{equation}
m\left\langle S\left(x\right) \right\rangle_{Q=1} = \chi_{s,m=0,V} m^2 + \frac{1}{V} 
\left(2 + O\left(m^2\right)\right)
\label{scondensate2fq1}
\end{equation}
which gives for the difference between the full 
condensates in the $Q=1$ and $Q=0$ topological 
sectors the following expression

\begin{equation}
m\left(\left\langle \sum_x\left(S_u\left(x\right)+ S_d\left(x\right)\right)
\right\rangle_{Q=1} - \left\langle \sum_x\left(S_u\left(x\right)+ 
S_d\left(x\right)\right)\right\rangle_{Q=0}\right)
= 2 + O\left(m^2\right)
\label{diffscondensate2f}
\end{equation}

\subsection{ Three or more flavors}
The generalization of the results of the previous subsection to $N_f\ge3$ is straightforward 
but with some peculiar features which we want to remark. 

In the two flavor model the scalar $\chi_{s,m=0,V}$ and pseudoscalar $\chi_{p,m=0,V}$ 
susceptibilities in the chiral limit get contributions from the $Q=0$ and $Q=1$ topological 
sectors. The $Q=0$ sector is free from the anomaly, and then gives the same contribution 
to both susceptibilities, but the $Q=1$ sector contributions to $\chi_{s,m=0,V}$ and 
$\chi_{p,m=0,V}$ are opposite. For $N_f\ge3$ however only the $Q=0$ sector gives contribution 
to the scalar $\chi_{s,m=0,V}$ and pseudoscalar $\chi_{p,m=0,V}$
susceptibilities in the chiral limit, and therefore we get 

\begin{equation}
\chi_{s,m=0,V} = \chi_{p,m=0,V},\hskip 0.4cm if\hskip 0.4cm N_f\ge3. 
\label{vaya}
\end{equation}

The expansion of the
mean value of the chiral condensate in powers of $m$ for $N_f=3$ 

\begin{equation}
\left\langle S\left(x\right) \right\rangle_\theta = \chi_{s,m=0,V} m -
\sin^2 \frac{\theta}{3}\left(\chi_{s,m=0,V} - \chi_{p,m=0,V}\right) m
+ O\left(m^2\right)
\label{scondensate3f}
\end{equation}
is therefore $\theta$-independent at order $m$, its first $\theta$-dependent contribution 
being of the order of $m^2$. In general the first $\theta$-dependent contribution to 
the expansion of 
the scalar condensate in powers of the quark mass $m$ is of the order of $m^{N_f-1}$, 
and therefore of the 
order of $m^{N_f}$ in the expansion of the vacuum energy density, analogous to equation 
(\ref{evac2f}). This is the reason why the generalization of equation 
(\ref{diffscondensate2f}) to $N_f$ flavors reads now as follows

\begin{equation}
m\left(\left\langle \sum_{f,x} S_f\left(x\right)
\right\rangle_{Q=1} - \left\langle \sum_{f,x} S_f\left(x\right)\right\rangle_{Q=0}\right)
= N_f + O\left(m\right)
\label{diffscondensateNf}
\end{equation}

\section{Summary and Conclusions}
The axion mass, an essential ingredient in the calculation of the axion abundance in the 
Universe, is related in the $QCD$ axion model with the topological susceptibility $\chi_T$ 
at $\theta=0$. The temperature dependence of the topological susceptibility is therefore 
just that of the axion mass. 

Three papers reporting numerical results for the temperature dependence of the topological 
susceptibility in unquenched $QCD$ have been recently published \cite{martinelli}, 
\cite{petre}, \cite{javier}, and their conclusions seem not to be in agreement 
with each other, 
reflecting the high level of difficulty in measuring the topological susceptibility in the 
high temperature regime.

In this paper we have shown that an analysis of some of the numerical results reported
in \cite{javier}, concerning the mean value of the chiral condensate at fixed topological
charge, suggests that the vacuum energy density is $\theta$-independent at high temperatures,
but surprisingly not too high $(T\sim 2T_c)$, a result which would imply a vanishing
topological susceptibility, and the absence of all physical effects of the $U(1)$ axial
anomaly at these temperatures. More precisely we have shown that 
the results of the numerical simulations of $QCD$ at $T>T_c$ in \cite{javier}, \cite{frison},
are in contradiction with the results of the large volume expansion around the saddle 
point (\ref{difactions}), (\ref{difcondensates}), but in very good agreement with the 
analytical perturbative expansion of the chiral condensate given by equation 
(\ref{diffscondensateNf}).

The only reason for the failure of the saddle point expansion is that the
main ingredient of this expansion, the non-trivial $\theta$-dependence of the vacuum energy 
density $E(\beta, m_f, \theta)$, does not work. Other intermediate solutions, like a 
vacuum energy density with non trivial $\theta$ dependence for $|\theta|\le\theta_c$, which 
becomes $\theta$-independent at $|\theta|>\theta_c$, would imply the existence of a phase 
transition at $(T, \theta_c)$, and such a situation seems to be ruled out if $T\ge T_c$, 
at least in the pure gauge model, by the results of \cite{delia}, \cite{new}, 
which show that the critical temperature of the deconfinement phase transition decreases 
with $\theta$. Therefore a $\theta$-independent 
vacuum energy density seems the most plausible explanation for the failure of 
the saddle point expansion. 

In section 4 we have made a general assumption concerning the high temperature phase of 
$QCD$, where the $SU(N_f)_A$ symmetry is restored. Basically we assume that in this phase 
all correlation
lengths are finite in physical units, none symmetry is spontaneously broken, the model
is free from infrared divergences, and the
perturbative expansion of the chiral condensate in powers of the quark mass converges. A
finite spatial lattice volume of linear size much larger than the inverse mass gap should
be enough to reproduce the correct physical results, and contrary to what happens in the
low temperature broken phase, the infinite volume limit and the chiral limit should commute.
Under this assumption we have shown that all effects of the axial
anomaly should disappear in the high temperature phase of $QCD$, where the
$SU(N_f)_A$ symmetry is restored. 
The topological susceptibility and all the $\theta$-derivatives of the vacuum energy density
should vanish and the theory should become $\theta$-independent at any $T>T_c$. 

Incidentally, the commutativity of the chiral and infinite volume limits is 
implicitly assumed by the authors of reference \cite{javier}, since their 
calculations in the high temperature regime are based on the assumption that 
the topological susceptibility, $\chi_T$, can be computed in this phase from the 
relation $V\chi_T = \frac{2Z_1}{Z_0}$, which is just the result that follows from 
the dilute instanton gas approximation in the $V\chi_T\ll 1$ limit.

An analysis of the physical implications of these results on the axion cosmology seems 
therefore worthwhile.

The author thanks Giuseppe Di Carlo, Eduardo Follana and Alejandro Vaquero for very
long and deep discussions on the topological properties of QCD, and Javier Redondo
for very useful discussions on the results of reference \cite{javier}. The author also
thanks the referee of this paper for his criticisms and comments.
This work was funded by MINECO under grant FPA2015-65745-P (MINECO/FEDER).

\vfill
\eject


\begin{thebibliography}{99}
\bibitem{peccei}
R.D. Peccei,
\textit{AIP Conf.Proc. 1274} (2010) 7.


\bibitem{weinberg}
S. Weinberg,F. Wilczek, 
\textit{Physical Review Letters} \textbf{40}, (1978) 223.

\bibitem{wilczek}
F. Wilczek, 
\textit{Phys. Rev. Lett.} \textbf{40}, (1978) 279.

\bibitem{pq}
R.D. Peccei, H.R. Quinn, 
\textit{Phys. Rev. Lett.} \textbf{38}, (1977) 1440; {Phys. Rev.} \textbf{D16}, (1977) 1791.

\bibitem{turner}
M. Turner, 
\textit{Physical Review} \textbf{D33}, (1986) 889.

\bibitem{vicari}
E. Vicari, H. Panagopoulos, 
\textit{Phys. Rep.} \textbf{470}, (2009) 93.

\bibitem{pv}
H. Panagopoulos, E. Vicari, 
\textit{JHEP} \textbf{1111}, (2011) 119.

\bibitem{bdpv}
C. Bonati, M. D'Elia, H. Panagopoulos, E. Vicari, 
\textit{Phys. Rev. Lett.} \textbf{110}, (2013) 252003.

\bibitem{new}
C. Bonati, M. D’Elia and A. Scapellato, 
\textit{Phys. Rev. D} \textbf{93}, (2016) 025028.

\bibitem{enrico}
E. Berkowitz, M.I. Buchoff, and E. Rinaldi
\textit{Phys. Rev. D} \textbf{92}, (2015) 034507.

\bibitem{japo}
R. Kitano and N. Yamada, 
\textit{JHEP} \textbf{1510}, (2015) 136.

\bibitem{javier_quen}
Sz. Borsanyi et al., 
\textit{Phys. Lett.} \textbf{B752}, (2016) 175.

\bibitem{martinelli}
C. Bonati, M. D’Elia, M. Mariti, G. Martinelli, M. Mesiti, F. Negro, F. Sanfilippo,
G. Villadoro, 
\textit{JHEP} \textbf{1603}, (2016) 155.

\bibitem{petre}
P. Petreczky, H-P. Schadler and S. Sharma, 
\textit{arXiv:1606.03145 [hep-lat]} (2016).

\bibitem{javier}
Sz. Borsanyi et al., \textit{Nature} \textbf{ 539} (2016) no.7627, 69-71, 
\textit{arXiv:1606.07494 [hep-lat]} (2016).

\bibitem{cohen1}
T. D. Cohen, 
\textit{Phys. Rev. D} \textbf{54}, (1996) 1867.

\bibitem{cohen2}
T. D. Cohen, 
\textit{nucl-th/9801061} (1998).

\bibitem{cuatro}
C. W. Bernard et al., 
\textit{Physical Review Letters} \textbf{78}, (1997) 598.

\bibitem{cinco}
S. Chandrasekharan, D. Chen, N. H. Christ, W. -J. Lee, R. Mawhinney and P. M. Vranas,
\textit{Phys. Rev. Lett.} \textbf{82}, (1999) 2463.

\bibitem{seis}
H. Ohno, U. M. Heller, F. Karsch and S. Mukherjee, 
\textit{PoS LATTICE 2011} (2011) 210.

\bibitem{siete}
A. Bazavov et al., 
\textit{Physical Review} \textbf{D86}, (2012) 094503.

\bibitem{ocho}
T. G. Kovacs and F. Pittler, 
\textit{PoS LATTICE 2011} (2011) 213.

\bibitem{nueve}
G. Cossu, S. Aoki, S. Hashimoto, T. Kaneko, H. Matsufuru, J. -i. Noaki and E. Shintani, 
\textit{PoS LATTICE 2011} (2011) 188.

\bibitem{diez}
S. Aoki, H. Fukaya and Y. Taniguchi, 
\textit{Phys. Rev. D} \textbf{86}, (2012) 114512.

\bibitem{once}
G. Cossu, S. Aoki, H. Fukaya, S. Hashimoto, T. Kaneko and H. Matsufuru, 
\textit{Physical Review } \textbf{D87}, (2013) 114514; Erratum: 
\textit{Physical Review } \textbf{D88},
(2013) 019901.

\bibitem{doce}
G. Cossu et al.,
\textit{PoS LATTICE 2015} (2016) 196.

\bibitem{docebis}
B.B. Brandt, A. Francis, H.B. Meyer, O. Philipsen, D. Robainad and H. Wittig,
\textit{arXiv:1608.06882 [hep-lat]} (2016).

\bibitem{wolfgang}
R. Brower, S. Chandrasekharan, J.W. Negele and U.-J. Wiese,
\textit{Phys. Lett. B} \textbf{560}, (2003) 64.

\bibitem{aokiqfix}
S. Aoki, H. Fukaya, S. Hashimoto, and T. Onogi,
\textit{Phys. Rev. D} \textbf{76}, (2007) 054508.

\bibitem{Smilga}
H. Leutwyler and A. Smilga,
\textit{Phys. Rev. D} \textbf{46}, (1992) 5607.

\bibitem{Ginsparg}
P.H. Ginsparg and K. G. Wilson,
\textit{Phys. Rev. D} \textbf{25}, (1982) 2649.

\bibitem{Neuberger}
H. Neuberger,
\textit{Phys. Lett. B}\textbf{417} (1998) 141;
\textit{Phys. Lett. B}\textbf{427} (1998) 353.

\bibitem{Luscher}
M. Luscher, \textit{Phys. Lett. B} \textbf{428} (1998) 342.

\bibitem{Victor}
P. Hasenfratz, V. Laliena and F. Niedermayer,
\textit{Phys. Lett. B} \textbf{427}, (1998) 125 [{\tt hep-lat/9801021}].

\bibitem{monos}
V. Azcoiti, G. Di Carlo, E. Follana and A. Vaquero, 
\textit{JHEP} \textbf{1007}, (2010) 047.

\bibitem{frison}
J. Frison, R. Kitano, H. Matsufuru, S. Mori, and N. Yamada,
\textit{arXiv:1606.07175 [hep-lat]} (2016).

\bibitem{delia}
M. D'Elia and F. Negro,
\textit{Phys. Rev. Lett.} \textbf{109}, (2012) 072001;
\textit{Phys. Rev. D} \textbf{88}, (2013) 034503.




\end{thebibliography}
\end{document}